\documentclass[%
reprint,
superscriptaddress,
preprintnumbers,
 amsmath,amssymb,
 aps,
prd,
]{revtex4-2}

\usepackage{graphicx}% Include figure files
\usepackage{dcolumn}% Align table columns on decimal point
\usepackage{bm}% bold math
\usepackage{physics}
\usepackage{slashed}
\usepackage{floatrow}
\usepackage{comment}

\usepackage[utf8]{inputenc}
\usepackage{multirow}
\usepackage{url}
\usepackage{color}
\usepackage{natbib}
\usepackage{hyperref}
\usepackage{enumitem}
\usepackage{ulem}
\usepackage{times,txfonts}
\usepackage[T1]{fontenc}
\usepackage{ae,aecompl}

\newcommand{\e}{\mathrm{e}}
\newcommand{\iu}{\mathrm{i}}
\newcommand{\g}[1]{\gamma^{#1}}

\newcommand{\us}[4]{u_{#1,#2}^{\left(#3\right)}\left(#4\right)}

\newcommand{\vs}[4]{v_{#1,#2}^{\left(#3\right)}\left(#4\right)}

\newcommand{\lb}{\lambda_B^2}

\renewcommand{\Vec}[1]{\ensuremath{\bm{\mathrm{#1}}}}

\renewcommand{\vec}[1]{ \bm{#1}} % works for roman and greek letters
\newcommand{\nvec}[1]{ \hat{\bm{#1}}}

\newcommand{\prlsection}[1]{\section{#1}}
\newcommand{\prlsubsection}[1]{\subsection{#1}}

\begin{document}

\preprint{HIP-2026-7/TH}

\title{QED cross sections in strong magnetic fields}

\author{Olavi Kiuru}
\email{olavi.kiuru@helsinki.fi}
\affiliation{Department of Physics and Helsinki Institute of Physics,
P.O.~Box 64, FI-00014 University of Helsinki, Finland}
\author{Joonas Nättilä}
\email{joonas.nattila@helsinki.fi}
\affiliation{Department of Physics and Helsinki Institute of Physics,
P.O.~Box 64, FI-00014 University of Helsinki, Finland}
\author{Risto Paatelainen}
\email{risto.paatelainen@helsinki.fi}
\affiliation{Department of Physics and Astronomy, FI-20014 University of Turku, Finland}
\affiliation{Department of Physics and Helsinki Institute of Physics,
P.O.~Box 64, FI-00014 University of Helsinki, Finland}
\author{Aleksi Vuorinen}
\email{aleksi.vuorinen@helsinki.fi}
\affiliation{Department of Physics and Helsinki Institute of Physics,
P.O.~Box 64, FI-00014 University of Helsinki, Finland}

\date{\today}

\begin{abstract}

The magnetospheres of magnetars, a class of highly magnetized neutron stars, host magnetic fields exceeding the Schwinger limit, where Quantum Electrodynamics (QED) becomes nonlinear. 
In such environments, QED scattering processes are strongly modified, which may affect plasma dynamics. 
In this work, we apply a formalism originally developed for the study of magnetic-field effects in hot quark–gluon plasma to strong-field QED. 
The method resums interactions between virtual electrons and the external magnetic field, consistently incorporating the finite decay widths of excited Landau levels derived from the fermion self-energy.
Using this framework, we perform the first systematic analysis of tree-level QED scattering processes in strong magnetic fields, concentrating on the processes of highest relevance for the plasma dynamics of magnetars.
All resulting cross sections are provided in an open-source Python package.

\end{abstract}

\maketitle

\prlsection{\label{sec:intro}Introduction}
Quantum Electrodynamics (QED) is often referred to as the most accurately tested theory of physics, with theoretical predictions for quantities such as electron and muon anomalous magnetic moments, the fine structure constant, and various scattering cross sections agreeing with measurements to astonishing precision \cite{Aoyama:2020ynm}.
This is, however, only the case for vacuum QED, while the presence of a thermal medium or strong external fields qualitatively alter the situation.
In the latter context, dubbed strong-field QED (SFQED), interactions between charged particles and the background field $\mathcal{A}^\mu$ may turn QED nonlinear, giving rise to new physical effects, such as photon self-interactions, and complicating perturbative calculations even at low orders.

Among physical systems, where electromagnetic fields reach values high enough to probe the nonlinear regime of QED, the two most widely studied ones include high-intensity lasers \citep{fedotov_advances_2023, gonoskov_charged_2022} and the magnetospheres of a special class of neutron stars (NSs) called magnetars \citep{duncan_formation_1992, kaspi_magnetars_2017, rea_magnetars_2025}. 
In the latter context, the background magnetic fields can exceed $10^{15} \,\mathrm{G}$, reaching tens of times the Schwinger field  $B_Q \equiv m_e^2/e \approx 4.41\cdot 10^{13}\;$G (with $c=\hbar=1$), where the electron's cyclotron energy $eB/m_e$ becomes comparable to its rest mass $m_e$. 
In this limit, the background field induces qualitative effects upon physical quantities, such as scattering cross sections, which must be accounted for when describing the collective dynamics of the system.

\begin{figure}
    \includegraphics[width=0.95\textwidth]{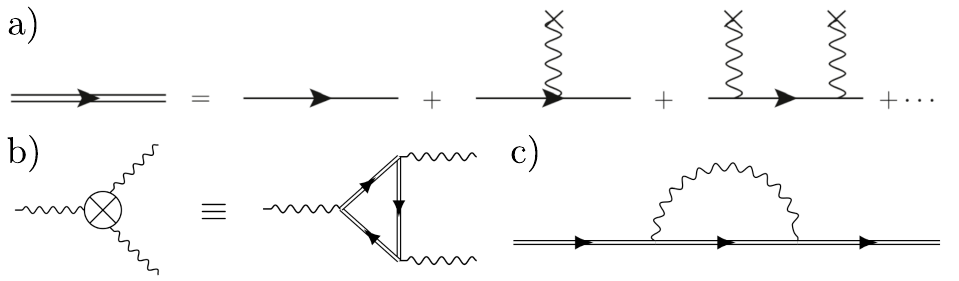}
    \caption{a) In the Furry expansion, the fermion propagator includes all possible interactions with $\mathcal{A}^\mu$, denoted here by squiggly lines that end with $\cross$ \citep{fedotov_advances_2023}. b) In a background magnetic field, the 3-photon fermion loop is non-zero and yields an effective 3-photon vertex. Together with the Furry expansion, this photon self-interaction gives rise to the nonlinear effects of SFQED. c) The fermion self-energy $\Sigma$ regulates some divergences present in SFQED cross sections.}
    \label{fig:furryPropagator}
\end{figure}

In the presence of a strong background field, it is useful to split the gauge field in two parts in the QED Lagrangian. 
Treating the background field $\mathcal{A}^\mu$ as part of the noninteracting Lagrangian $\mathcal{L}_0$ but leaving the dynamical gauge field $A^\mu$ in the interaction part $\mathcal{L}_\mathrm{int}$, we write \footnote{Note that we assume $\mathcal{A}^\mu$ to satisfy Maxwell's equations, so that it does not affect the kinetic term of the gauge field in $\mathcal{L}_0$.}
\begin{eqnarray}
    \mathcal{L} &=& \mathcal{L}_0 + \mathcal{L}_\mathrm{int}, \\
    \mathcal{L}_0&\equiv& -\frac{1}{4}F_{\mu\nu}F^{\mu\nu} + \Bar{\psi}\left(\iu\slashed{\partial} - q \slashed{\mathcal{A}} -m \right)\psi, \;\, \mathcal{L}_\mathrm{int} \,\equiv\, -q\Bar{\psi}\slashed{A}\psi.\;\;\;\; \nonumber
\end{eqnarray}
A fermion propagator derived from $\mathcal{L}_0$ captures the background-field effects to all orders, while a well-behaved weak-coupling expansion in $e$ or $\alpha_e\equiv e^2/(4\pi)$ can be derived using $\mathcal{L}_\mathrm{int}$.
This rearrangement constitutes the well-known Furry picture of QED \citep{furry_bound_1951}, illustrated in Fig.~\ref{fig:furryPropagator}.

For an electron in a background magnetic field, $\vec{B}=B\nvec{z}$ (and $\vec{E}=\vec{0}$), the available energy levels are functions of the electron mass, its momentum component along the magnetic field $p^z$, and the magnetic scale $\lambda_B^{-1}\equiv\sqrt{eB}$ that is associated with the quantized Landau levels related to the electron's transverse motion, 
\begin{equation}\label{eq:LandauLevels}
E^2=p_z^2+m_e^2+2neB, \quad n\in \mathbb{N} .
\end{equation} 
The need for the Furry description can be seen from here:
as soon as the scale $\lambda_B^{-1}$ becomes comparable to $m_e$ and $p_z$, i.e.~$b\equiv B/B_Q=eB/m_e^2\sim 1$, the magnetic field begins to qualitative modify the energy spectrum of a free electron.

Magnetar magnetospheres are studied by numerically simulating the strongly magnetized relativistic plasma, with microphysical input parametrized through the SFQED cross sections.
While their derivation has been extensively studied in the literature, important caveats remain. 
Existing studies typically focus on just a single process, e.g., Compton scattering \citep{mushtukov_compton_2016}, or make assumptions that severely limit the applicability of the results, such as working in the Lowest-Landau-Level (LLL) limit for all (external and virtual) fermions \citep{kostenko_qed_2018, kostenko_qed_2019}. 

In this work and its extended companion paper \citep{companion}, we determine the cross sections of all tree-level 1-to-2, 2-to-1, and 2-to-2 QED scattering processes with no internal photon lines; cf.~Table \ref{tab:processes}.
We do this using a formalism originally developed for the study of magnetic catalysis in  heavy-ion collisions \cite{shovkovy_magnetic_2013, miransky_quantum_2015}, which features resummed fermion propagators, complete sums over Landau levels, and the associated spectral widths.
This way, the effects of the background field become consistently accounted for, extending the applicability of the results to all values of the magnetic field.
Finally, we note that all final expressions and their numerical implementations are provided in an open-source Python package \footnote{To appear at \url{https://github.com/hel-astro-lab}.}.

\begin{table}[]
    \caption{The QED scattering processes considered in this work. In the column "Status", N stands for a new result, V for the verification of an existing one, and F for a process left for the future, while the column "Rate" indicates the scaling of the expected rate of interactions with the number densities of $e^\pm$ and photons. Given that both of the processes marked with F scale with powers of $n_\pm$ (and $n_\pm \ll n_\gamma$), their effect on magnetosphere plasma dynamics is expected to be suppressed.}\label{tab:processes}
    \begin{tabular}{lccc}
    \hline \hline
    \multicolumn{2}{c}{Process} & Status & Rate \\
    \hline \hline
    Synchrotron radiation & $e^\pm \to e^\pm + \gamma$ & V & $n_\pm$ \\
    One-$\gamma$ pair creation & $\gamma \to \e^+ + e^-$ & V & $n_\gamma$ \\
    One-$\gamma$ pair annihilation & $e^+ + e^- \to \gamma$ & N & $n_\pm^2$ \\
    Synchrotron self-absorption & $e^\pm + \gamma \to e^\pm$ & N & $n_\pm n_\gamma$ \\
    \hline
    Compton scattering & $e^\pm + \gamma \to e^\pm + \gamma$ & V & $n_\pm n_\gamma$ \\
    Two-$\gamma$ pair creation & $\gamma + \gamma \to e^+ + e^-$ & N & $n_\gamma^2$ \\
    Two-$\gamma$ pair annihilation & $e^+ + e^- \to \gamma + \gamma$ & N & $n_\pm^2$ \\
    \hline
    M\o ller scattering & $e^\pm + e^\pm \to e^\pm + e^\pm$& F & $n_\pm^2$ \\
    Bhabha scattering & $e^+ + e^- \to e^+ + e^-$& F & $n_\pm^2$ \\
    \hline \hline
    \end{tabular}
\end{table}

\prlsection{\label{sec:setup} Setup}
We are interested in the tree-level differential cross sections of various QED scattering processes. In the Landau gauge, $\mathcal{A}^\mu = Bx\nvec{y}$, these can be written as
\begin{equation}\label{eq:crossSec}
\sigma = \int \prod_j \dd\left[\mathrm{PS}\right]_j (2\pi)^3 \delta^{(3)}\left(E,p^z,p^y\right)\frac{\abs{M_{\mathrm{fi}}}^2}{F},    
\end{equation}
where $j$ runs over all outgoing particles;
\begin{equation}
   \dd \left[\mathrm{PS}\right]_\gamma \equiv \frac{\dd[3]{k}}{2\omega(2\pi)^3}, \quad \dd\left[\mathrm{PS}\right]_{e^\pm} \equiv \frac{\dd{p^y}\dd{p^z}}{2E(2\pi)^2}
\end{equation}
denote the integration measures; $F$ is the initial state particle flux; the delta functions
\begin{equation}  \delta^{(3)}\left(E,p^z,p^y\right)\equiv \delta(E_\mathrm{in}-E_\mathrm{out})\delta(p^z_\mathrm{in}-p^z_\mathrm{out})\delta(p^y_\mathrm{in}-p^y_\mathrm{out})
\end{equation}
enforce the relevant conservation laws; and $\abs{M_{\mathrm{fi}}}^2$ is the squared matrix element of the scattering process, obtained using the relevant Feynman rules. 

If the initial state has one particle, the flux equals the number density of the incoming particles, ${F=n=2E_\mathrm{in}}$~\footnote{This nonstandard definition of the particle number density is due to a different choice of normalization of one-particle states. They are normalized such that there is $2E$ particles per unit volume. For more details see \citep{cannoni_lorentz_2017}.}, with $E_\mathrm{in}$ the corresponding energy, while for a two-particle initial state with densities $n$ and $n'$, ${F=nn'\sqrt{\left(\vec{v}-\vec{v}'\right)^2-\left(\vec{v}\cross\vec{v}'\right)^2}\equiv 4E_\mathrm{in} E_\mathrm{in}'\bar{v}}$ \citep{cannoni_lorentz_2017}. 
All components of Eq.~\eqref{eq:crossSec} are independently Lorentz invariant under boosts (anti-)parallel to $\vec{B}$ for both 2-to-1 and 2-to-2 scatterings. 
For 1-to-2 scattering, the cross section is to be understood as a decay rate and scales as the energy of the particle under Lorentz boosts. 
The differences in Eq.~\eqref{eq:crossSec} compared to vacuum QED are explained by the fact that the perpendicular components $p^x$ and $p^y$ of the $e^\pm$ momenta are gauge dependent and, in general, not conserved due to interactions with $\mathcal{A}^\mu$. 
In the Landau gauge, $p^x$ is not defined while $p^y$ is conserved and related to the center of gyration, $a \equiv \frac{q}{\abs{q}}\lb p^y$ ($q=+e$ for positrons and $-e$ for electrons). 
Importantly, $a$ is not an observable, and the obtained cross sections must be averaged over all possible values of $a$ for each incoming $e^\pm$, i.e. 
$\bar{\sigma} = \int \prod_l \left(\lb  \dd{p^y_l}\right) \sigma$ where $l$ runs over all incoming $e^\pm$, ensuring gauge independence.

The most nontrivial part of Eq.~\eqref{eq:crossSec} is clearly the matrix element $M_\mathrm{fi}$, which must be evaluated for each process using the formalism briefly outlined above. Below, we summarize the key elements of this computation, while additional technical details are given in Appendix \ref{app:formalism} and our companion paper \citep{companion}.

\vspace{0.1cm}

\prlsubsection{Fermion propagator}
The most notable difference to vacuum QED is visible in the fermion propagator, for which we use a mixed representation featuring a momentum-space Fourier transform in $t$ and $z$, but not in $x$ and $y$. 
The propagator includes all possible interactions with $\mathcal{A}^\mu$, as shown in Fig.~\ref{fig:furryPropagator} a., and features a sum over all Landau levels of the virtual fermion. The explicit expression of the propagator is given in Eq.~\eqref{eq:fermionProp} of Appendix \ref{app:formalism}. An advantage of this representation of the propagator is that it can be easily generalized to nonzero temperature, chemical potentials, or chiral shifts \citep{miransky_quantum_2015, ghosh_fermion_2024, wang_photon_2021}. 

\vspace{0.1cm}

\prlsubsection{Decay widths} 
The cross sections determined using Eq.~\eqref{eq:fermionProp} lead to unphysical resonances, i.e.~energies at which the cross section diverges due to the fermion propagator going on shell. To consistently regulate them, we first determine the imaginary part of the fermion self-energy $\Sigma$
and then include the result in the dressed propagator used in the evaluation of the matrix elements. A nonzero imaginary part morphs the divergences into resonance peaks of nonzero width determined by $\Im \Sigma$, while a coupling of the fermions to $\mathcal{A}^\mu$ can be seen to lift the spin degeneracy of the Landau levels \citep{geprags_electron_1994,gonthier_compton_2014}.

\citet{ghosh_fermion_2024} determined a one-loop result for $\Im \Sigma$ (Feynman diagram shown in Fig.~\ref{fig:furryPropagator} c.) at non-zero temperature using the formalism adopted here. We have confirmed a perfect match of the zero-temperature limit of this result with the synchrotron emission rate of an on-shell electron.
It is, however, worth noting that the resonances associated with the LLL are not damped by this effect at $T=0$ because an electron at LLL does not emit synchrotron radiation.

\vspace{0.1cm}

\noindent \prlsubsection{External fermions}
The plasma dynamics of magnetar magnetospheres is dominated by electrons, positrons and photons, as photon energies are expected to be low enough, so that heavier leptons are not produced in large amounts. 
In a classical description, an electron traveling in a background magnetic field gyrates around the magnetic field lines with the gyrofrequency ${\omega_B = e\abs{B}/m_e}$ (so that $\omega_B = m_e$ when $B=B_Q$), but is otherwise constrained to move along the magnetic field lines. 
As discussed around Eq.~\eqref{eq:LandauLevels} above, the corresponding energies become quantized, which can be interpreted either as the quantization of the perpendicular momentum or that of the gyration radius. 
It is worth noting that while higher Landau levels permit both spin states, all LLL electrons are spin down and all LLL positrons spin up, and that all Landau level energies are independent of $a$. 

If one is only interested in spin-averaged cross sections, there is no need to specify the spin states of the external fermions, but for spin-specific cross sections we need to choose a basis for the spinor wave functions. 
In astrophysics calculations, Sokolov-Ternov (ST) wave functions are  typically preferred due to their Lorentz transformation properties and the fact that their two spin states decay independently via synchrotron radiation \cite{sokolov_synchrotron_1968, gonthier_compton_2014, herold_cyclotron_1982, gonthier_compton_2000}. 
The electron and positron ST wave functions are given in Appendix \ref{app:formalism} Eqs.~\eqref{eq:ElectronST} and \eqref{eq:PositronST}, respectively. 

\vspace{0.1cm}

\noindent \prlsubsection{External photons}
Photons can be described by three quantities: their energy $\omega$, normalized three-momentum $\nvec{k}$, and polarization state. 
They propagate in two polarization modes: the ordinary and extraordinary, or O- and X-modes. 
In the O-mode, the electric field of the photon oscillates in a plane spanned by $\nvec{k}$ and $\vec{B}$, while in the X-mode the oscillations are perpendicular to this plane. 
As we will show below, the SFQED cross sections depend strongly on polarization as well as $k_\perp$, i.e., the perpendicular momentum of the photon with respect to $\vec{B}$, which stays invariant under Lorentz boosts parallel to $\vec{B}$.

Because thermal photons from the magnetar photosphere have energies $\omega \ll m_e$, we expect the corresponding quantum corrections to be small, so that we may treat photons similarly to the vacuum QED aside from the polarization states. 
In this approximation, we lose out on one important scattering process---photon splitting $\left(\gamma \to \gamma \gamma\right)$---which is forbidden in vacuum QED and originates entirely from quantum corrections \citep{adler_photon_1970, adler_photon_1971}. 
This process is important for magnetospheric plasma dynamics because unlike one-photon pair creation, it has no low-energy kinematic threshold and can therefore act as the generator of particle cascades even for $k_\perp < 2m_e$, the threshold energy for pair creation. 
Should quantum corrections to photons be found important, they can be derived from the vacuum polarization diagram \citep{wang_photon_2021}.

\vspace{0.1cm}

\noindent \prlsubsection{Rates and approximations}
Finally, we briefly comment on two issues related to the interpretation of our results: the rates referred to in Table \ref{tab:processes} and the approximations used here and in previous literature. 
The reaction rate $\mathcal{R}_\mathrm{f}$ is defined as the number of events to a specific final state per unit time and volume \citep{cannoni_lorentz_2017}, related to the cross section via $\mathcal{R}_\mathrm{f} = \sigma F$.
This quantity is proportional to the number densities of the incoming particles and, since $n_\gamma \gg n_{e^\pm}$ in the magnetosphere, we expect scattering processes including photons to dominate the dynamics.

\begin{figure}
    \centering
    \includegraphics[width=0.9\linewidth]{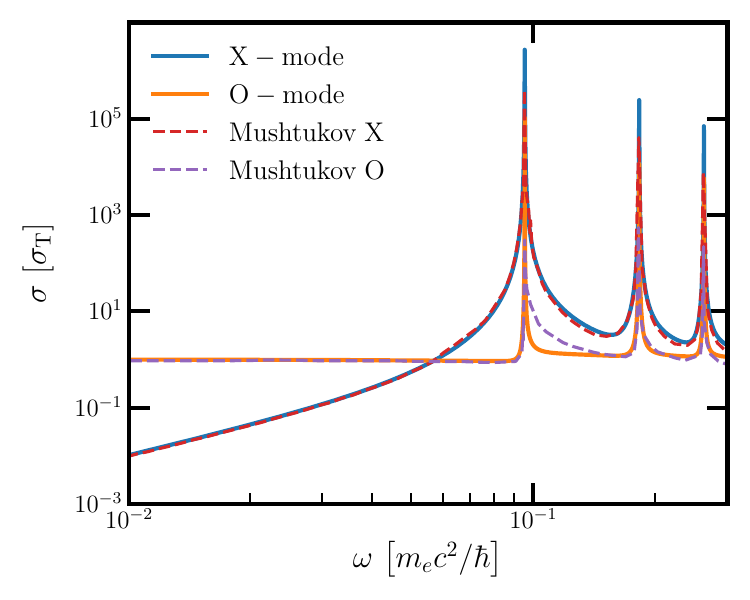}
    \caption{Total Compton cross section for a photon moving in a direction perpendicular to the magnetic field, with $b=0.1$. The blue and orange solid lines stand for our results for the X- and O-modes, respectively, while the corresponding results of Ref.~\citep{mushtukov_compton_2016} are denoted by red and violet dashed lines.}
    \label{fig:ComptonComparison}
\end{figure}

A common way to simplify cross-section calculations in SFQED is to use the LLL approximation (LLLa), where both the external and internal (virtual) fermions are restricted to the $n=0$ Landau level \citep{kostenko_qed_2018,kostenko_qed_2019}. 
Assuming that the incoming fermions are on the lowest Landau level is typically a reasonable assumption since, in a strong magnetic field, any fermion in an excited state will rapidly de-excite to the ground state by emitting a photon through synchrotron radiation. 
The exclusion of excited Landau levels in virtual and outgoing fermions is, on the other hand, far less motivated, but some approximations are nevertheless permitted. 
The Landau levels of outgoing fermions are limited by the total energy of the scattering, and while the Landau levels of virtual fermions are in principle unbounded, their contribution falls off rapidly when their number surpasses the sum of the Landau levels of external fermions \citep{kozlenkov_two-photon_1986}. 
While completely ignoring excited Landau levels is not justified, it is thus a valid approximation to only consider a finite number of them.

\prlsection{\label{sec:results}Results}
Moving next to a discussion of our main results, we note that we neither display all cross sections obtained nor list their analytic forms here due to their high number and at times extreme length. 
This information is instead provided through an open-source Python package that contains all results in a ready-to-use form and a longer companion paper that provides the related computational details \citep{companion}. The discussion in this section focuses on 2-to-2 particle scattering processes. For a brief discussion on one-photon pair annihilation, see Appendix \ref{app:1ppa}.

The scattering process that has received maximal attention in existing literature is Compton scattering, which we are thus able to use as a benchmark point. 
In Fig.~\ref{fig:ComptonComparison}, we display our result for the total Compton cross section of a photon traveling in a direction perpendicular to the magnetic field (solid lines) and  compare it with fig.~4.a.~of \citep{mushtukov_compton_2016} (dashed lines). 
The cross section is summed over outgoing spin and polarization states, taking into account all allowed Landau levels for the outgoing electron. 

The two results align perfectly until the first resonance, after which some minor differences are observed, of which the differing heights of the resonance peaks can be explained by the slightly different implementations of decay widths. 
In Ref.~\citep{mushtukov_compton_2016}, the decay widths were added as imaginary shifts in the energies of the external electrons as an attempt to model the finite lifetimes of higher Landau levels.  
In contrast, we treat Compton scattering and synchrotron radiation separately, so that they can be self-consistently implemented to QED Monte Carlo codes, such as that presented in \citep{nattila2024}. 
The other differences in the cross section, most visible in the O-mode, appear to be related to excited Landau levels of external fermion legs and are further inspected in Appendix \ref{app:comparison}.

\begin{figure}
    \centering
    \includegraphics[width=0.9\linewidth]{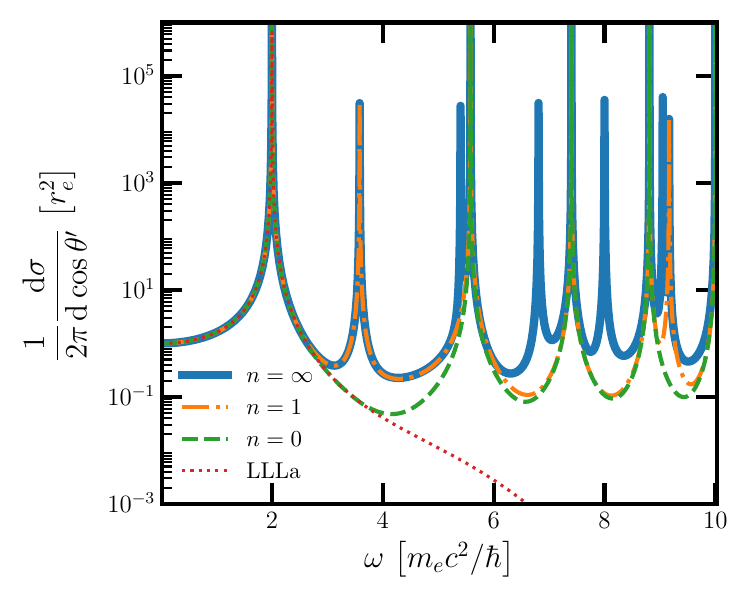}
    \caption{Differential cross section for Compton scattering, with both (O-mode) photons moving perpendicular to the magnetic field ($\theta=\flatfrac{\pi}{2}=\theta'$), $n_i=0$, and $b=10$. The solid blue line stands for the full result, while the other lines denote various approximations as explained in the main text.}
    \label{fig:Compton}
\end{figure}

In Fig.~\ref{fig:Compton}, we next investigate the validity of different approximation schemes by studying the  differential Compton cross section for a perpendicular collision, where the incoming fermion resides on the $n_i=0$ Landau level. 
We observe that the LLLa result described above (red dotted line) is valid for $\omega \lesssim 2.5 m_e$, but falls off sharply at higher photon energies, while a more accurate result is obtained when all Landau levels available for the outgoing electron are taken into account. 
If the propagator is nevertheless restricted to the LLL $n=0$ (green dashed line), several higher resonances still missed, but including the lowest excited Landau level in the propagator (orange dash-dotted line) improves the situation up to $\omega \approx 5m_e$. 
Beyond this, one needs to include multiple Landau levels in the propagator, culminating in the full result (solid blue line) where contributions from sufficiently many Landau Levels are included in the internal propagator.
Such a difference in the cross section dramatically modifies high-energy electron dynamics, as it makes the photon targets opaque while they remain unphysically transparent in the LLLa. 

As discussed above, the resonances visible in the LLLa results are not damped and are therefore infinite, while other resonances have finite size. 
The heights of these peaks can be approximated by $\sigma_{n,\mathrm{res}} \sim \big[(\Gamma^+_n+\Gamma^-_n)/2\big]^{-2}$, i.e., they are related to the spin-averaged decay widths. 
The region of validity of the LLLa grows with the strength of the magnetic field and, as discussed in \citep{kostenko_qed_2018}, can be extensive for very large magnetic fields. 
However, for the typical field strengths of magnetar magnetospheres $b\lesssim 30$, the approximation fails already at moderate frequencies.

\begin{figure}
    \centering
    \includegraphics[width=0.87\linewidth]{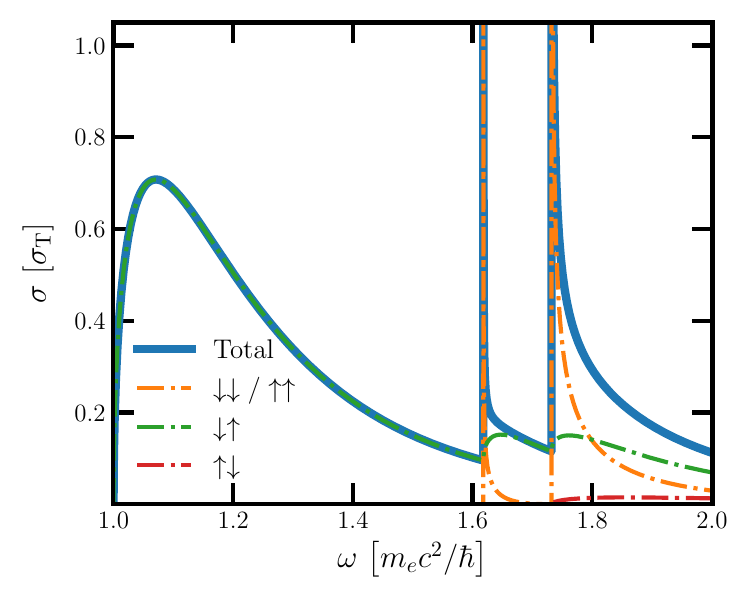}
    \caption{Two-photon-pair-creation cross section for a head-on collision of photons moving perpendicular to the magnetic field with $\omega=\omega'$ and $b=1$. The dash-dotted lines correspond to the production of specific spin states as indicated in the legend, with the left arrow always standing for the electron and the right one for the positron. The solid blue line shows the sum of the four combinations.}
    \label{fig:2PPC}
\end{figure}

The existing results for two-photon pair creation are based on the use of Johnson-Lippmann (JL) spinors \citep{kozlenkov_two-photon_1986} that do not allow for the extraction of spin-dependent cross sections \citep{gonthier_compton_2014}. 
The choice of spinors should not make a difference if the result is summed over all spin states, but as discussed in Appendix \ref{app:comparison}, our spin-summed results do not fully agree with Ref.~\citep{kozlenkov_two-photon_1986} even then.

In Fig.~\ref{fig:2PPC}, the novel spin dependence of our results is scrutinized for a head-on collision of photons moving perpendicular to the magnetic field lines. 
Given that only a spin-down state is available for a ground-state electron and similarly a spin-up state for a ground-state positron, these are the only spin states produced for small collision energies. 
This combination of spins is seen to dominate the cross section even for higher photon energies, except near the thresholds of new scattering channels.

\prlsection{\label{sec:discussion}Discussion}
In this work, we have presented a systematic framework for determining QED scattering cross sections in the background of ultrastrong magnetic fields, relying on machinery originally developed for the study of magnetic catalysis in hot quark-gluon plasma \cite{shovkovy_magnetic_2013,miransky_quantum_2015}. 
Using this scheme, we have addressed all tree-level processes of relevance for the plasma dynamics of neutron-star magnetospheres, with results showcased here and more extensively covered in the accompanying Python package \footnotemark[2] and companion paper \citep{companion}.

The most central object in our formalism is the resummed fermion propagator, which treats interactions with the background field to all orders; cf.~Appendix \ref{app:formalism} Eq.~\eqref{eq:fermionProp}. 
This result---as our entire computational setup---can be straightforwardly generalized to the presence of a nonzero temperature, higher orders in both loops and external legs, and different types of background fields. 
An example of the latter can be found in laser-plasma physics, where the background is a linearly polarized electromagnetic wave and experiments will soon probe the non-linear regime of QED \cite{fedotov_advances_2023, gonoskov_charged_2022}. 
The closest astrophysical parallel would involve an electric field parallel to the magnetic one, leading to the $z$-dependence of spinors being replaced by parabolic cylinder functions \cite{nikishov_equivalent_2003}. 

To consistently model the non-linear dynamics of radiative plasmas in magnetar magnetospheres, one needs detailed information about individual SFQED processes, obtainable from the cross sections presented in this work. 
The particle number densities $n_\gamma$ and $n_\pm$ are set by pair creation and annihilation processes, while Compton scattering and pair creation are responsible for equilibrating the particle distributions and for creating fermions on excited Landau levels. 
In strong magnetic fields, the excited Landau levels mostly decay to the LLL through synchrotron radiation, producing high-energy photons capable of pair creation into excited Landau levels. 
This chain reaction can continue for multiple cycles, leading to an exponential production of $e^\pm$-pairs, known as pair cascades. 
Some of the photons created during a pair cascade can also escape the magnetosphere and contribute to the observed hard X-ray emission of magnetars. 

Existing magnetospheric simulations typically employ vacuum cross sections (e.g., \citep{zeng_origin_2025, beloborodov_electron-positron_2013}), leading to interaction rates that may be orders-of-magnitude too small and giving rise to qualitatively incorrect plasma dynamics.
Some recent simulations (e.g., \citep{zhang_quantum_2025}) have employed cross sections evaluated using LLLa, which, however, also fails for high-energy particles $p_\perp \gtrsim m_e\sqrt{b}$, for which excited Landau levels dominate the cross sections.

To remedy for these shortcomings, our open-source Python library captures contributions from all Landau levels and grants the user access to all degrees of freedom, including Landau levels, polarization, spin, magnetic field strength, and particle momenta. 
This enables building efficient pipelines to simulation codes, allowing the user to investigate how different processes affect plasma dynamics and paving the way to a more detailed understanding of magnetar emission spectra \citep{kaspi_magnetars_2017}.

\prlsection{Acknowledgments}
This work has been supported by the European Research Council through the ERC StG grant ILLUMINATOR (101114623); the Research Council of Finland (RCF) through projects 354533 and 347499 as well as  the Centre of Excellence in Neutron-Star Physics (projects 374062 and 374063); and the Finnish Cultural Foundation.

\bibliography{apssamp}

\appendix

\section{Feynman rules}\label{app:formalism}

%In the End Matter, we provide additional  details central for the derivation of the results discussed above, and in addition display selected additional figures complementing those presented in the main text.

\prlsubsection{Landau-gauge fermion propagator}
We work in the Landau gauge, performing a Fourier transform in the $t$- and $z$-coordinates, but not in $x$ or $y$. In this case, the fermion propagator we employ reads
\begin{eqnarray} \label{eq:fermionProp}
    S_\mathrm{F}(x',x;E, p^z) &\equiv& \iu \frac{\e^{\iu\Phi(x',x)}\e^{-\frac{\xi^2}{2}}}{2\pi\lb} \sum_{n=0}^\infty \frac{F_n}{E^2-(p^z)^2-m^2-2n\abs{qB}}, \nonumber \\
\end{eqnarray}
with the various objects defined as
\begin{eqnarray}    &&F_n=\;\left(\g{0}E-\g{3}p^z+m\right)\left( \mathcal{P}_{-}L_{n-1}(\xi^2) + \mathcal{P}_{+}L_n(\xi^2) \right) \nonumber \\ 
    &&+ \frac{\iu}{\lb} \Vec{\gamma}_\perp\vdot(\Vec{x}_\perp-\Vec{x}'_\perp)L_{n-1}^1(\xi^2), \; \quad \mathcal{P}_\pm \equiv \frac{1}{2}\left(1\pm \frac{q}{|q|}\iu \g{1}\g{2}\right), \nonumber \\
    &&\xi^2 \equiv \frac{(x-x')^2 + (y-y')^2}{2\lb}, \:\quad
    \Phi(x',x) \equiv - \frac{q}{|q|} \frac{(x+x')(y-y')}{2\lb}, \nonumber
\end{eqnarray}
and $L_n$ and $L_n^{(\alpha)}$ standing for ordinary and generalized Laguerre polynomials. Finally, we note that the insertion of the fermion self-energy to this propagator is discussed around eqs.~(C1)--(C3) of the companion paper \citep{companion}.

\prlsubsection{Spinors}
The ST spinors we employ take the forms
\begin{equation}
    \psi^{(\sigma)}_{n,a}(x^\mu) = \begin{cases}
        & \us{n}{a}{\sigma}{x}\e^{-\iu p\cdot x} \text{ (electrons)} \\
        & \vs{n}{a}{\sigma}{x}\e^{\iu p\cdot x} \text{ (positrons)}
    \end{cases},
\end{equation}
where
\begin{equation}\label{eq:ElectronST}
\begin{split}
    \us{n}{a}{-1}{x} &= \frac{1}{f_n}\mqty(\mqty{-\iu p^z p_n \phi_{n-1,a}(x)  \\ (E+E_{0})(E_{0}+m)\phi_{n,a}(x) \\ -\iu p_n (E+E_{0})\phi_{n-1,a}(x) \\ -p^z(E_{0}+m)\phi_{n,a}(x)}), \\
    \us{n}{a}{+1}{x} &= \frac{1}{f_n}\mqty(\mqty{(E+E_{0})(E_{0} + m)\phi_{n-1,a}(x)  \\ -\iu p^z p_n\phi_{n,a}(x) \\ p^z(E_{0}+m)\phi_{n-1,a}(x) \\ \iu p_n (E+E_{0})\phi_{n,a}(x)}),
\end{split}
\end{equation}
and
\begin{equation}\label{eq:PositronST}
\begin{split}
    v_{n,a}^{(+1)}(x) &= \frac{1}{f_n}\mqty(\mqty{- p_n (E+E_{0})\phi_{n-1,a}(x) \\ -\iu p^z(E_{0}+m)\phi_{n,a}(x) \\ -p^z p_n \phi_{n-1,a}(x) \\ \iu (E+E_{0})(E_{0}+m)\phi_{n,a}(x)}), \\
    v_{n,a}^{(-1)}(x) &= \frac{1}{f_n}\mqty(\mqty{-\iu p^z(E_{0}+m)\phi_{n-1,a}(x) \\ -p_n (E+E_{0})\phi_{n,a}(x) \\ -\iu (E+E_{0})(E_{0} + m)\phi_{n-1,a}(x)  \\ p^z p_n\phi_{n,a}(x)}).
\end{split}
\end{equation}
Here, we have employed the short-hand notation
\begin{equation}
    \begin{split}
        p_n^2 &\equiv 2n\abs{qB} \equiv E_{0}^2-m^2, \\
        f_n &\equiv \sqrt{2E_{0}(E_{0}+m)(E_{0}+E)}, \\
        \phi_{n,a}(\Vec{x})&\equiv \frac{1}{\left(\pi^\frac{1}{2}\lambda_B 2^n n!\right)^\frac{1}{2}}H_n\left(\frac{x-a}{\lambda_B}\right)\e^{-\frac{\left(x-a\right)^2}{2\lb}},
    \end{split}
\end{equation}
with $H_n$ denoting Hermite polynomials, and note that in the Landau gauge $p^\mu = (E, 0, p^y, p^z).$
In a given Feynman diagram, an incoming electron or positron corresponds to the expression $\us{n}{a}{\sigma}{x}\e^{-\iu p\cdot x}$ or $\vs{n}{a}{\sigma}{x}^\dagger\g{0}\e^{-\iu p\cdot x}$, respectively, while an outgoing electron or positron corresponds to $\us{n}{a}{\sigma}{x}^\dagger\g{0}\e^{\iu p\cdot x}$ or $\vs{n}{a}{\sigma}{x}\e^{\iu p\cdot x}$.

\vspace{0.1cm}

\prlsubsection{Photons}
Analogously to the external fermion lines, an incoming photon with four-momentum $k^\mu$ corresponds to the expression $A^\mu (x) = \varepsilon^\mu \e^{\iu k\cdot x}$
and an outgoing photon to $\left(A^\mu (x)\right)^*$, while the polarization vectors of the O- and X-modes read
\begin{equation}
    \varepsilon_\mathrm{O}^z=\sin{\theta}, \quad\varepsilon_\mathrm{O}^\pm=\varepsilon_\mathrm{O}^x\pm\iu\varepsilon_\mathrm{O}^y=-\cos{\theta}\e^{\pm\iu\phi}
\end{equation}
and 
\begin{equation}
    \varepsilon_\mathrm{X}^z=0,\quad\varepsilon_\mathrm{X}^\pm=\mp\iu\e^{\pm\iu\phi},
\end{equation}
respectively. 
The photon propagator finally remains unchanged from vacuum QED, taking the Feynman-gauge form
\begin{equation}
    D_{\mu\nu} = \frac{-\iu g_{\mu\nu}}{p^2}.
\end{equation}

\vspace{0.1cm}

\prlsubsection{Photon-fermion vertex} The photon-fermion vertex also remains unchanged from vacuum QED and reads
\begin{equation}
    \Gamma^\mu = -\iu q \g{\mu} \int \dd[4]{x}.
\end{equation}

\vspace{0.1cm}

\prlsubsection{Decay widths} As discussed in the main text, the inclusion of decay widths in the evaluation of SFQED cross sections is needed to regulate divergencies and produce physically meaningful results. 
In astrophysical computations, the decay widths have traditionally been obtained from the synchrotron decay rate \citep{herold_cyclotron_1982}, while the quantum-field-theoretically consistent way of taming the resonances proceeds through the calculation of the fermion self-energy, given by the Feynman diagram in Fig.~\ref{fig:furryPropagator}.c.

\begin{figure}
    \centering
    \includegraphics[width=0.9\linewidth]{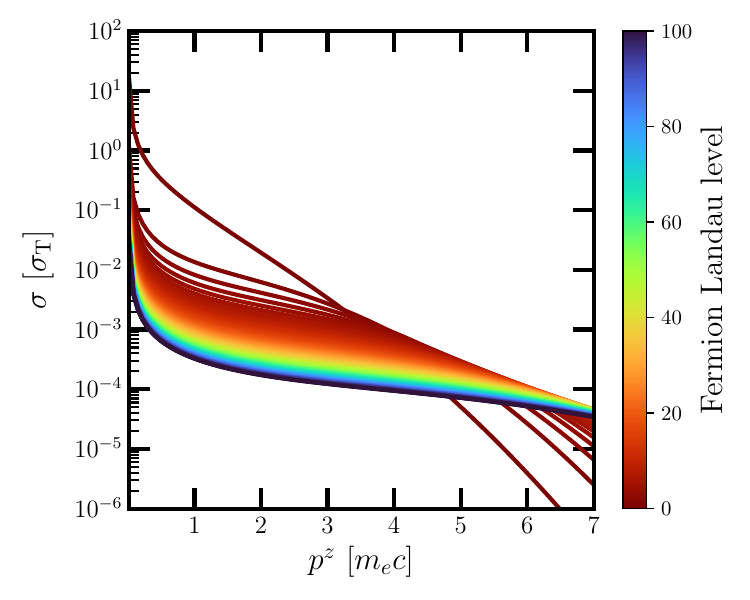}
    \caption{Cross section of one-photon pair annihilation as a function of the electron longitudinal momentum $p^z$ at $b=10$, averaged over incoming spin states and summed over outgoing polarization states. The color coding refers to the shared Landau levels of the electron and positron, which we vary from 0 to 100.}
    \label{fig:1PPA}
\end{figure}

The connection between the two methods has been discussed in, e.g., Ref.~\citep{graziani_strong-field_1993}, and for tree-level scattering, there should be no meaningful difference. 
We have confirmed this by determining the decay width from the fermion self energy and confirming that it indeed matches the synchrotron result.

\prlsection{Pair annihilation}\label{app:1ppa}
In all previous studies of QED cross sections at strong magnetic fields, the incoming fermions have been assumed to reside on the LLL due to the rapid decay of excited Landau levels. 
This, however, does not account for relativistic effects extending the lifetimes of these states, which makes excited Landau levels of potential interest for simulations involving ultrarelativistic fermions. 
Given that the formalism we use allows for the incoming fermions to be placed on arbitrary Landau levels, in Fig.~\ref{fig:1PPA} we study the dependence of the one-photon pair-annihilation cross section on the Landau levels of the incoming electron-positron pair, working in a frame where the total momentum in the $z$ direction vanishes. 
We observe a non-monotonous behavior, where the cross section is maximized for both fermions residing on the lowest Landau level when $p^z\lesssim 3m_e$, but higher Landau levels dominate at larger momenta.

\prlsection{Comparison to earlier results}\label{app:comparison}

\begin{figure}
    \centering
    \includegraphics[width=0.9\linewidth]{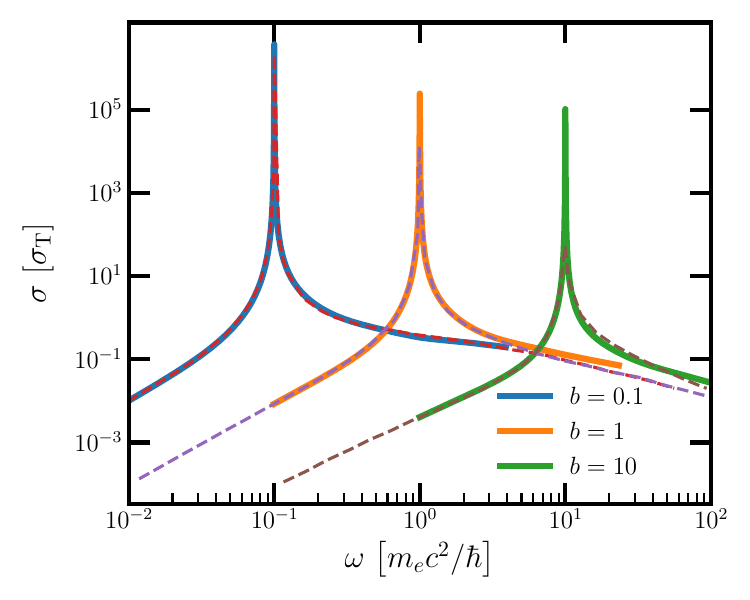}
    \caption{Total cross section of Compton scattering for an O-mode photon moving parallel to the magnetic field for different magnetic field strengths. The solid lines show our results, while the dashed lines show the results obtained by \citet{mushtukov_compton_2016}, fig.~3.}
    \label{fig:ComptonParallelComparison}
\end{figure}

In Fig.~\ref{fig:ComptonParallelComparison}, we next compare our results for the Compton cross section to the results obtained in Ref.~\citep{mushtukov_compton_2016}. Inspecting photons moving along the magnetic field (solid lines), we find a good match between the two results, except for slight deviations at large energies, where the excited Landau levels of the outgoing electron play an increasingly crucial role in the cross section \footnote{Note, that there is a typo in eq.~(7) of Ref.~\citep{mushtukov_compton_2016}; in the denominator there should be $s_f$ instead of $s_f^2$.}. Also, as discussed in the Results section, the heights of the resonance peaks are somewhat different compared to our cross sections due to different implementations of the decay widths. We have also confirmed that our analytical results match with the ones obtained in Ref.~\citep{daugherty_compton_1986} aside from a few minus signs \citep{companion}.

\begin{figure}
    \centering
    \includegraphics[width=0.9\linewidth]{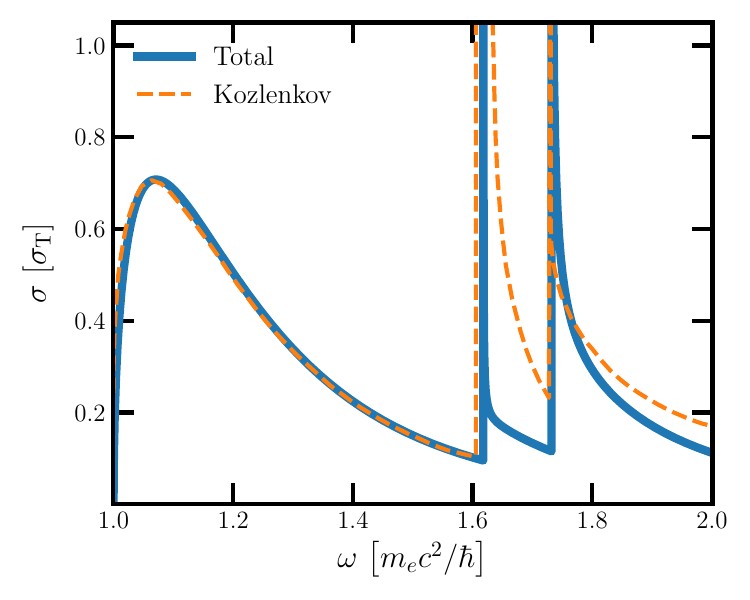}
    \caption{Two-photon pair creation cross section summed over fermion spin states for the same setup considered in Fig.~\ref{fig:2PPC}. The solid line corresponds to our result while the dashed one displays the result obtained by \citet{kozlenkov_two-photon_1986}, fig.~4.}
    \label{fig:2ppcComparison}
\end{figure}

In Fig.~\ref{fig:2ppcComparison}, we finally compare our result for the spin-summed cross section of two-photon pair creation (solid line) to the one obtained in Ref.~\citep{kozlenkov_two-photon_1986}. The cross section has two infinite peaks due to the opening of new scattering channels. The first peak corresponds to the case where one of the fermions is on the LLL and the other one is on the second excited Landau level, while the second peak stems from both fermions residing the first excited Landau level. We see that our results agree with Kozlenkov and Mitrofanov before the first peak, i.e., in the region where the external fermions are on the LLL, but differ in the region where the excited Landau levels of the external fermions need to be taken into account.

\end{document}